\documentstyle[prl,aps,multicol,epsfig,tabularx]{revtex} 

\begin{document} 
 
\draft 
 
\title{Segregation transitions in wet granular matter} 
\author{Azadeh Samadani and A. Kudrolli} 
\address{Department of Physics, Clark University, Worcester, MA 
01610} 
 
\date{\today} \maketitle 

\begin{abstract} 
We report the effect of interstitial fluid on the extent of segregation by imaging the pile that results after bidisperse color-coded particles are poured into a silo. Segregation is sharply reduced and preferential clumping of small particles is observed when a small volume fraction of fluid $V_f$ is added. We find that viscous forces in addition to capillary forces have an important effect on the extent of segregation $s$ and the angle of repose $\theta$. We show that the sharp initial change and the subsequent saturation in $s$ and $\theta$ occurs over similar $V_f$. We also find that a transition back to segregation can occur when the particles are completely immersed in a fluid at low viscosities. 
\end{abstract}

\pacs{PACS number(s): 45.70.Mg, 47.50.+d, 81.05.Rm} 

\begin{multicols}{2}

The flow of granular matter shows a number of fascinating properties including segregation. Segregation in granular matter can be a benefit or a nuisance depending on the application. Remarkable sensitivity to small differences in the size, shape and density of the constituent grains has been observed~\cite{bridgwater93,rosato87,zik94,makse97,jaeger96,shinbrot00}. However, most of this work relates to noncohesive granular matter. Although decrease of percolation in granular matter immersed in a fluid has been noted~\cite{bridgwater78}, systematic studies are required. Furthermore, quantitative experiments on segregation have not been reported in the situation where a small amount of fluid or humidity is present.

Presence of small amounts of interstitial fluid in the system introduces another degree of complexity due to the cohesive forces between particles in addition to the core repulsion force and the friction force present for dry granular matter. Increase in the angle of repose is the most well known effect of presence of interstitial fluid in a granular system and has been a topic of current interest~\cite{hornbaker97,tegzes99,halsey98,bocquet98,fraysse99}. The interstitial fluid also alters the percolation of particles, and the particles tend to behave as clumps rather than individual grains. 

The first quantitative study of the segregation of bidisperse granular matter in the presence of interstitial fluid is reported. The extent of segregation is parameterized by visualizing color-coded particles of different sizes that are poured into a quasi-two dimensional silo. A sharp reduction of segregation is observed when a small volume fraction of fluid $V_f$ is added which introduces capillary bridges between particles. Preferential clumping of small particles is observed to cause layering at small $V_f$. We obtain the segregation phase diagram as a function of size ratio $r$ of the bidisperse particles and $V_f$. We show the importance of the viscous force in addition to the capillary force on both the extent of segregation and the angle of repose $\theta$ by changing the viscosity $\nu$ of the fluid. We find that the sharp change in the extent of segregation and $\theta$ occurs over similar $V_f$. The development of segregation is also measured when the particles are completely immersed in the fluid. The extent of segregation decreases with viscosity $\nu$ of the fluid and disappears for large enough $\nu$. This decrease is also observed to depend on $r$.

%\section{Experimental Apparatus} 

A rectangular silo of dimensions $50.0\, {\rm cm} \times 30.5\, {\rm cm}$ and a width $w$ of $3.0\,{\rm cm}$ is used for the experiments. The flow is visualized through the glass side-walls of the silo using a megapixel Kodak ES 1.0 digital camera. The glass beads and fluids used are listed in Table~\ref{fluids}. The granular sample is prepared in batches by throughly mixing 1.0~kilograms each of the two kinds of beads with a volume fraction $V_f$ of the fluid before the granular material is placed inside the reservoir. The volume fraction $V_f$ corresponds to the ratio of the volume of the fluid to the volume of all the particles. The material is first filled into a reservoir and then drained through a pipe into the silo. A tall cylindrical reservoir is used to minimize segregation from occuring during pouring~\cite{arteaga90,samadani99}. Pipes with different diameters are used to control the flow rate $Q$. Wider diameters are used at higher $V_f$ in order to maintain $Q \sim 2.2$~g/s. The reservoir is raised at a slow constant rate with a stepper motor and a system of pulleys. The slow upward velocity of the reservoir allows the particles to accumulate inside the pipe before flowing down the surface and reduces the kinetic energy of particles due to free fall. Thus the interaction of particles is restricted to that of a moving surface layer with the static bulk. The flow is confined to a thin surface layer and is continuous at $V_f = 0$ but becomes slip-stick as $V_f$ is increased.

To minimize the effect of evaporation, the experiments are conducted within 
a few minutes after the sample is mixed. Measurement with glycerol and 
polybutene does not show systematic effects of evaporation by delaying the 
time between mixing and pouring from 3 to 30 minutes. The data for water at 
low $V_f$ ( $< 2 \times 10^{-3}$), shows some effects of evaporation for long delay 
times.

%\section{Observations}

\begin{figure} 

\begin{center} 

\centerline{\epsfig{file=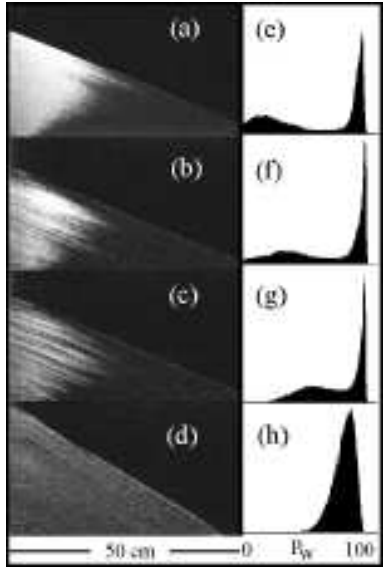,width=5 cm}} 

\end{center} 

\caption{(a)-(d) Images of the resulting piles after the granular mixture is slowly poured into the silo from a reservoir ($r = 2.4$). The volume fraction of water (a) $V_f = 0 $, (b) $V_f = 6\times 10^{-4}$, (c) $V_f = 1.2 \times 10^{-3}$, and (d) $ V_f = 6 \times 10^{-3}$. The segregation decreases rapidly with increase in $V_f$. Aperiodic layering is observed under certain conditions (c). (e)-(h) The histogram of local particle density ratio $P_w$ of the two kinds of particles in the pile. The peaks in the distribution can be fitted to Gaussians and the mean value used to determine the most common particle ratios $a$ and $b$. } 

\label{seg-par-def} 

\end{figure}

Fig.~\ref{seg-par-def}(a) represents the segregation displayed by bidisperse dry granular matter flowing down an inclined plane. The segregation mechanism in the dry case is well known and can be explained using (i) the void filling mechanism where smaller particles percolate through the larger particles and are thus found at the bottom of the flow~\cite{lun88}, and (ii) the capture mechanism where the smaller particles which are more sensitive to surface fluctuations are stopped at the top of the pile before the larger particles~\cite{boutreux96}. The segregation is observed to vanish by adding a small amount of fluid  which corresponds to less than 1\% of the volume (see Fig.~\ref{seg-par-def}(d)). Fig.~\ref{seg-par-def}(b),(c) show the intermediate situation where partial segregation is observed. Some layering also occurs although it is not as periodic as in mixtures of particles with {\em dry} rough and smooth particles~\cite{makse97}.

To parameterize the extent of segregation, histograms of the ratios of the two types of particles in a $3.5\,{\rm mm^2}$ area are made using the light intensity. The light intensity is a monotonic function of the density ratio of the particles and is determined by using known weight ratios of particles in a separate series of calibration experiments. The histogram of local particle density ratio $P_w$ of the two kinds of particles in the pile is thus obtained and is plotted in Fig.~\ref{seg-par-def}(e)-(h). The peaks in the distribution can be fitted to Gaussians and the mean value used to determine the most common particle ratios $a$ and $b$.  The segregation parameter is defined as $s= \frac{(b-a)}{100}$ which has a value between 0 and 1 and describes the extent of segregation.

\begin{figure} 

\begin{center} 

\centerline{\epsfig{file=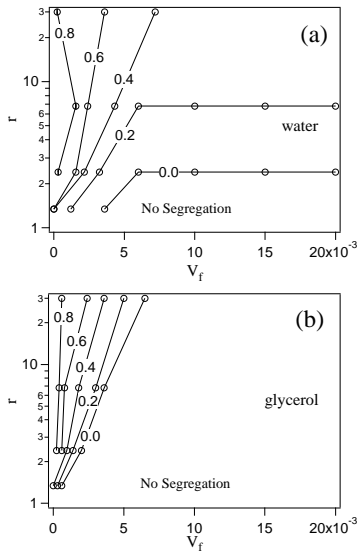,width=5 cm}} 

\end{center}

\caption{(a) Phase diagram of the segregation as a function of $r$ and $V_f$ of water. (b) Phase diagram for glycerol. Systematic data is obtained up to $V_f = 25 \times 10^{-3}$ but a slightly lower range is shown. Since the segregation-mixing transitions are continuous and also in some cases not complete, contour lines are used to denote the iso-$s$ at each $r$ where a measurement was made. The two fluids have different $\nu$ but similar $\Gamma$, see Table~\ref{fluids}.  Segregation is observed at low $V_f$ but decreases sharply especially for low $r$ and higher $\nu$.}

\label{phase} 

\end{figure}

The phase diagram of segregation as a function of $r$ and $V_f$ is plotted in Fig.~\ref{phase} for water and glycerol. The $s=0$ contour line is observed to close up at much lower $V_f$ for glycerol which is almost a thousand times more viscous than water but has similar surface tension $\Gamma$. The phase diagram not only demonstrates the importance of $r$ and $V_f$ on determining the extent of segregation but also $\nu$. The phase diagram also shows that segregation can persist even in the presence of cohesive forces for large $r$.

\begin{figure} 

\begin{center} 

\centerline{\epsfig{file=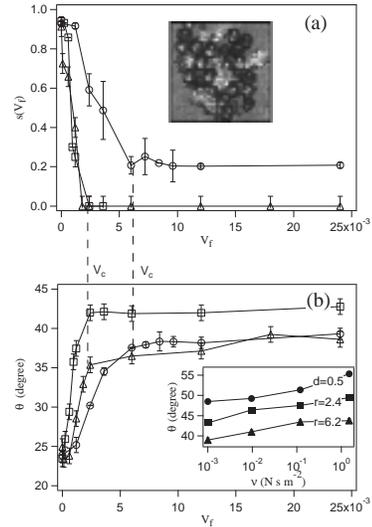,width=5 cm}} 

\end{center}

\caption{(a) The $s(V_f)$ as a function of $V_f$ at $r = 6.2$ for water ($\circ$), glycerol ($\Box$) and polybutene L-50 ($\triangle$). Inset: Clumps of small particles are observed at low $V_f$ which behave as a single large particle with a rough surface. (b) $\theta$ as a function of $V_f$ for the corresponding $s(V_f)$. Note that the saturation in $s(V_f)$ and $\theta$ occur at similar $V_f$ and is indicated by the vertical dashed lines. The saturation occurs at different values $V_c$ depending on $r$, $\nu$ and $\Gamma$.   Inset: $\theta$ versus $\nu$ for monodisperse beads with $d=0.5$ mm, and bi-mixtures with $r = 2.4$ and $r = 6.2$ at fixed $V_f = 24 \times 10^{-3}$. The viscosity in this case is varied by mixing water and glycerol.}

\label{seg-par} 

\end{figure}

The $s$ for $r=6.2$ is plotted in Fig.~\ref{seg-par}(a) as a function of $V_f$ for three fluids with different $\nu$ and $\Gamma$. The corresponding change in angle of repose $\theta$ as a function of $V_f$ is plotted in Fig.~\ref{seg-par}(b) and indicates the increased cohesiveness of the material. The angle of repose increases linearly at low $V_f$ and saturates at high $V_f$ thus sharing qualitative features with the data reported in Refs.\cite{hornbaker97,tegzes99,halsey98,bocquet98,fraysse99}. In 
Refs.~\cite{halsey98,bocquet98,fraysse99}, the maximum angle of stability or 
repose was measured and estimated by tilting a silo filled with grains. In 
Ref.~\cite{tegzes99}, the $\theta$ was measured using a draining crater 
method. However, quantitative agreement is not observed and may not be 
surprising because of the differences in the definition of the angle of 
repose in all the studies. Indeed, the published results are not in mutual agreement. In addition, it can be noted from Fig.~\ref{seg-par} that $\theta$ increases rapidly as the $\nu$ of the fluid is increased. To show the dependence on $\nu$ more explicitly, the $\theta$ for monodisperse particles, BD-2, and BD-3 are plotted at fixed $V_f$ in the inset to Fig.~\ref{seg-par}(b). The $\nu$ is varied by mixing water and glycerol which have similar $\Gamma$~\cite{crc}. $\theta$ is observed to increase for all cases. It can be also noted from this plot that the $\theta$  depends on the size ratio $r$ of the grains and saturates at a different value for bi-mixtures than for one of its components. This dependence will be reported in greater detail in a future publication~\cite{samadani00b}.

It can be noted from Fig.~\ref{seg-par} that $s$ and $\theta$ change rapidly over similar $V_f$. The $V_f$ at which the $s$ and the $\theta$ begins to saturate is denoted by $V_c$ (dashed lines in Fig.~\ref{seg-par}.) Although the increase of cohesivity between the grains due to the presence of interstitial fluid is well known, the fact that $s$ and $\theta$ show similar dependence on $V_f$ is remarkable. 
This is also the case not only for water and glycerol but also for polybutene L-50 which has a different combination of $\Gamma$ and $\nu$. We find that $s$ and $\theta$ at other $r$ also show a $V_c$ which is lower for smaller $r$ and higher for larger $r$ and can be inferred from Fig.~\ref{phase}. 

Although there is a lack of theoretical and experimental consensus on the nature of the force at low $V_f$, the saturation force due to the surface tension at large $V_f$ is well known~\cite{hornbaker97,halsey98,bocquet98} and equals $\pi\Gamma d$ where $d$ is the diameter of the particle. For the particles ($d \sim 0.5$ mm) used in the experiments, and with glycerol or water as interstitial fluid, this force corresponds to $10^{-4}$ N. However, Figs.~\ref{phase} and~\ref{seg-par} indicate that $\nu$ has an effect on the segregation of granular material and on $\theta$ as well. 

To see how the viscous force is relevant, one considers a particular case of two particles in contact with a small amount of fluid in between. The importance of such a force has been discussed in the context of lubrication and friction~\cite{garnick99}. If $h$ is the thickness of the fluid between the particles, and $R$ is the radius of the circular contact, then the viscous force is approximately given by $\frac{3\pi}{2 h^3}\nu R^4\frac {\partial h}{\partial t}$ for $h\ll R$~\cite{persson98}. The curvature of the beads near the contact point is neglected. This formula states that the force directly depends on the rate of separation of the two particles $\frac {\partial h}{\partial t}$ and the viscosity of the fluid. 

If $\frac {\partial h}{\partial t}$ is assumed to be of the same order as the average flow velocity of the particles down the surface then $\frac {\partial h}{\partial t} \approx$ 1 mm/s, and for $h \approx 10\,\mu$m and $R \approx 100\,\mu$m, the viscous force between particles can be estimated to be $5 \times 10^{-4}$ N, about 5 times larger than the saturation capillary forces for glycerol. On the other hand for water which has lower $\nu$, the viscous force will be at least 100 times smaller and therefore irrelevant. As one increases $V_f$, the formula for the simple situation considered will need to be modified especially to take into account the shearing motion as well as the separation. 

The viscous force damps velocity fluctuations as it increases with relative velocity between particles. Since velocity  fluctuations and percolation among particles is required for segregation, $s$ is therefore observed to be lower at higher $\nu$ for similar $\Gamma$. 

In addition to the overall cohesiveness of the granular matter introduced by the fluid, an interesting effect observed below $V_c$ is the selective coating and clumping of smaller particles (see inset to Fig.~\ref{seg-par}(a)). A clump of small particles effectively behaves as a particle with a rough surface which can be larger than the individual large particle. Makse {\em et al.} have shown that stratification can occur when two species with different surface roughness are present in granular flows when the size of the rough particles exceeds the smoother particles~\cite{makse97}. The features observed in Fig.~\ref{seg-par-def}(c) appear to be related to this mechanism, although the increased stick-slip nature of the flow at the surface brought about by the addition of the fluid makes the layering aperiodic.  The strongest layering is observed for BD-2 (see Table I) and decreases at higher $r$ where the clumps of small particles stay smaller than the larger bead. The average size of clumps increases at higher $V_f$ and the interaction between small and large particles also becomes important. Thus clumps with small and large particles are formed resulting in smaller $s$.

%\subsection{viscous medium}

\begin{figure} 

\begin{center} 

\centerline{\epsfig{file=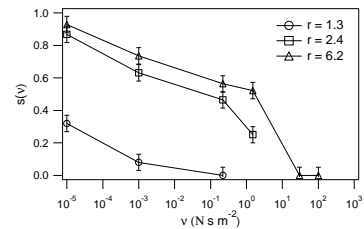,width=5 cm}} 

\end{center} 

\caption{ The extent of segregation $s(\nu)$ as a function of $\nu$ of the fluid in which particles of ratio $r$ are immersed is observed to decrease. The points at the lowest $\nu$ corresponds to air and $\nu$ is increased by using fluids that are listed in Table~\ref{fluids}.} 

\label{viscosity} 

\end{figure}

Finally the progress of segregation in the limit where particles are completely immersed in fluids was measured. In this case liquid bridges and capillary forces are absent but viscous forces are present. The experimental procedure and system is similar to that described for experiments with partial fluid fraction but with one major difference. The fluid with various $\nu$ is first filled into the silo, and then the mixture of dry bidisperse mixture is poured into the silo. Since the terminal velocity for particles decreases with $\nu$, the particles take a significantly longer time to drain in a viscous fluid compared to that in air and $Q < 0.02$g/s in glycerol. The $s$ is extracted after analyzing the images of the piles as discussed before and is plotted in Fig.~\ref{viscosity} as a function of $\nu$ for three different $r$. The segregation is observed to decrease as a function of $\nu$ and drops to zero at high $\nu$ which depends on $r$. In contrast to the case of partial fluid fraction, the angle of repose $\theta$ of the pile is not observed to change significantly with $\nu$.

The boundary layer which develops in the fluid for particles falling with velocity $v$ is proportional to $\sqrt{\nu d/\rho v}$ from dimensional arguments~\cite{batchelor} and washes away the details of the surface roughness of the pile at higher $\nu$. Thus the capture mechanism which is sensitive to surface roughness decreases in importance with $\nu$.  In addition, velocity fluctuations are damped out at higher $\nu$ as particles reach terminal velocity over a short distance leading to a reduction in percolation of particles~\cite{bridgwater78}. Thus, the segregation decreases because the two mechanisms which cause segregation in dry granular matter diminish. Although the terminal velocity reached by a spherical particle is proportional to $\sqrt{d}$,  this does not appear to aid segregation under the given conditions.

In summary, a sharp reduction of segregation is observed in granular flow when a small volume fraction of fluid is added. A transition back to segregation is observed when the volume fraction is increased so that particles are completely immersed in the fluid depending on the viscosity of the fluid. The role of the size ratio of the particles, volume fraction, surface tension and viscosity on the extent of segregation is clarified based on experiments and physical arguments. The results reported also offer a guide in the interpretation of  models of segregation developed for dry granular matter and geophysical phenomena where the presence of interstitial fluid is a fact. An interesting question which remains to be resolved is whether the segregation phase transitions change  qualitatively when the size of the system is increased arbitrarily and will be addressed in future work.

This work was supported by the National Science Foundation under grant number DMR-9983659 and by the donors of the Petroleum Research Fund.  A.K. also thanks the Alfred P. Sloan Foundation for its support.

%\end{multicols}

\end{multicols}

\begin{table}\begin{center} \begin{tabular}{l c c c c r c} \hline  

Bi-mixture \, & \, $d$ small (mm) \, & \, $d$ large (mm) \,& \,\,\,\,\,\,\,\,\,\,\,\, $r$ \,\,\,\,\,\,\,\,\,\,\,\, & \ $\theta$

\, \\ \hline \hline 

BD-1  &0.9  $\pm$ 0.1	&1.2 $\pm$ 0.1  	& 1.34 	&$23.0^\circ\pm 0.5^\circ$ \\

BD-2  &0.5 $\pm$ 0.1 	&1.2 $\pm$ 0.1	& 2.40 	&$23.0^\circ\pm 0.5^\circ$ \\ 

BD-3  &0.5 $\pm$ 0.1 	&3.1 $\pm$ 0.2 	& 6.20	&$25.0^\circ\pm 0.5^\circ$\\

BD-4  &0.1 $\pm$ 0.1 	&3.1 $\pm$ 0.2   	& 31.00     &$24.0^\circ\pm 0.5^\circ$

\\

\hline \end{tabular} \end{center} 

\begin{center} \begin{tabular}{l c c c} \hline  

Fluid \,  	& \, $ \nu $ $({\rm N s m^{-2}}) $ \, 	& \, $\rho$ $({\rm kg m^{-3}})$\, 	& \, $\Gamma$ $({\rm N m^{-1}})$\,\\ \hline \hline 

Water    		&0.001  		& 997		&0.073  \\

Glycerol 		&1.5  		&1126 	&0.070  \\ 

Polybutene L-50   &0.22		 	& 844  	&0.030  \\

Polybutene H-300  &\,\,\,\,\,30 @ 50$^0$ C	& 892		&-----  \\

Polybutene H-300	&100 			& 892		&-----  \\

\hline \end{tabular} \end{center} 

\caption{The bidisperse mixtures of spherical particles and fluids used in the experiments. $d$ is the diameter of the particles, $r$ is the size ratio, and $\theta$ is the angle of repose of the resulting (dry) pile. $\nu$ is the viscosity, $\rho$ the density and $\Gamma$ is the surface tension of the fluid. All data corresponds to 25$^0$ C unless stated otherwise.} 

\label{fluids} 

\end{table}

\end{document}